# From Microscale Variations to Macroscopic Effects: Directional Actuation, Phase Transition, and Negative Compressibility in Microfiber-Based Shape-Morphing Networks

*Shiran Ziv Sharabani, Elad Livnat, Maia Abuchalja, Noa Haphiloni, Nicole Edelstein-Pardo, Tomer Reuveni, Maya Molco, Amit Sitt\**

**Author**

    **Shiran Ziv Sharabani** - School of Chemistry, Faculty of Exact Sciences, the Center for Nanoscience and Nanotechnology, the Center for Physics & Chemistry of Living Systems, Tel Aviv University, Tel Aviv 6997801, Israel.

    **Elad Livnat** - School of Chemistry, Faculty of Exact Sciences, the Center for Nanoscience and Nanotechnology, the Center for Physics & Chemistry of Living Systems, Tel Aviv University, Tel Aviv 6997801, Israel.

    **Maia Abuchalja** - School of Chemistry, Faculty of Exact Sciences, Tel Aviv University, Tel Aviv 6997801, Israel.

    **Noa Haphiloni** - School of Chemistry, Faculty of Exact Sciences Tel Aviv University, Tel Aviv 6997801, Israel.

    **Nicole Edelstein-Pardo** - School of Chemistry, Faculty of Exact Sciences, the Center for Nanoscience and Nanotechnology, the Center for Physics & Chemistry of Living Systems, Tel Aviv University, Tel Aviv 6997801, Israel.

    **Tomer Reuveni** - School of Chemistry, Faculty of Exact Sciences, the Center for Nanoscience and Nanotechnology, the Center for Physics & Chemistry of Living Systems, Tel Aviv University, Tel Aviv 6997801, Israel.

    **Maya Molco** - School of Chemistry, Faculty of Exact Sciences, the Center for Nanoscience and Nanotechnology, the Center for Physics & Chemistry of Living Systems, Tel Aviv University, Tel Aviv 6997801, Israel.

**Corresponding Author**

    **Amit Sitt\*** – School of Chemistry, Faculty of Exact Sciences, the Center for Nanoscience and Nanotechnology, the Center for Physics & Chemistry of Living Systems, Tel Aviv University, Tel Aviv 6997801, Israel. Email: <u>amitsitt@tauex.tau.ac.il</u>

ABSTRACT: Two-dimensional shape-morphing networks are common in biological systems and have garnered attention due to their nontrivial physical properties that emanate from their cellular nature. Here, we present the fabrication and characterization of inhomogeneous shape-morphing networks composed of thermoresponsive microfibers. By strategically positioning fibers with varying responses, we construct networks that exhibit directional actuation. The individual segments within the network display either a linear extension or buckling upon swelling, depending on their radius and length, and the transition between these morphing behaviors resembles Landau's second-order phase transition. The microscale variations in morphing behaviors are translated into observable macroscopic effects, wherein regions undergoing linear expansion retain their shape upon swelling, whereas buckled regions demonstrate negative compressibility and shrink. Manipulating the macroscale morphing by adjusting the properties of the fibrous microsegments offers a means to modulate and program morphing with mesoscale precision and unlocks novel opportunities for developing programmable microscale soft robotics and actuators.

**TOC GRAPHICS**

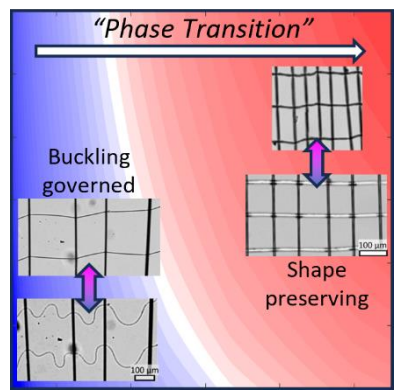

Developing shape-morphing materials with inhomogeneous and directional morphing is a longstanding objective in material programming, with extensive applications in soft robotics and bioinspired engineering.[1–4] Among various morphing systems, two-dimensional (2D) films have been widely studied for the formation of directional morphing, owing to the precise control that can be exerted over their physical and chemical attributes, as well as the ease of visualizing different morphing effects.[5–8] In recent years, there has been a growing interest in the mechanics of static and dynamic (quasi) 2D meshes and cellular architectures. These are characterized by an array of voids encapsulated by solid walls and offer advantages including a high surface area-to-volume ratio, low mass-to-volume ratio, and mechanical robustness, which made them highly suitable for design and fabrication using additive manufacturing approaches.[9–11] Moreover, cellular architectures can be employed to design mechanical metamaterials with unconventional mechanical properties such as auxetic behaviors and negative compressibility.[12,13]

The porous nature of the cellular architecture can introduce mechanical properties and morphing behaviors that significantly differ from continuous 2D sheets.[14,15] In addition, the cellular architecture provides inherent means to create anisotropy in the mechanical properties. The most trivial means is to impart distinct mechanical properties to the cellular walls.[16,17] For example, Mueller *et al.* demonstrated that constructing the cells' walls of two components, namely active and passive materials, arranged in a triangular network helped achieve anisotropic behavior under compression testing.[18] Anisotropy can also be induced by imposing an anisotropic cavity shape along the sample.[19]

While macroscopic shape morphing cellular structures are typically formed using 3D and 4D printing,[20,21] these approaches cannot be usually used for constructing microscale systems, where precise fabrication of the morphing components must be achieved. A common approach for obtaining shape-morphing cellular structures and meshes with microscale precision is lithography.[22] For example, soft lithography was used to create capillary-driven cellular structures that change shape and topology,[23] and a combination of surface deposition and lithography was employed to construct micro-lattices of 2D bonded films programmed to form curved 3D surfaces.[24]

Hierarchical networks of soft nano-to-microscale filaments represent another category of microscale shape-morphing cellular structures. Such networks are commonly encountered at the cellular level in biological systems (e.g., microtubules-kinesin and actin-myosin networks), and are used for exerting forces and inducing motility at the microscale level. A common approach for constructing synthetic shape-morphing networks of microscale filaments is through the spinning of microfibers of stimuli-responsive polymers.[25–27] Recently, highly structured networks of microfibers are obtained using the jet-writing approach, in which the spinning is coupled to a collector mounted on an accurate XY stage, enabling the deposition of ordered networks with high control over the network's architecture.[28–32] For example, Moon *et al.* employed jet-writing for the formation of polymer fiber meshes, which exhibited well-defined unidirectional morphing upon heating.[28] Javadzadeh *et al.* used melt-spinning jet-writing to form shape-morphing liquid-crystal elastomers scaffolds.[33] The transition to the microscale can affect the morphing behavior of the entire system at the macroscale. In previous work, we utilized the jet-writing approach to fabricate hierarchical networks of thermoresponsive mesoscale filaments that demonstrated two types of morphing: shape-preserving behavior and buckling-governed behavior.[29]

In this work, we present the development and characterization of responsive Cartesian networks composed of perpendicularly arranged thermoresponsive and non-responsive polymeric microfibers, which exhibit a directional and reversible morphing imposed by the inhomogeneity within the network's architecture. By changing the mesh size and consequently the slenderness of the filaments, we observe a transition in the morphing behavior at the single filament level between buckling-governed and linear extension regimes. Based on Landau's theory, we demonstrate that the transition can be regarded as an athermal second-order phase transition, where the mechanical energy of the system upon swelling is expressed as a function of a buckling-related order parameter, and the reciprocal slenderness serves as an effective temperature. Lastly, we demonstrate the ability to achieve both extensile and contractile regimes within a single network by changing the slenderness of the responsive fibers. This capability gives rise to a phenomenon analogous to "negative compressibility" and showcases the distinctive mechanical and shape-morphing characteristics exhibited by these networks.

A directionally morphing network was fabricated using the dry-spinning jet-writing approach. The network consisted of two-dimensional (2D) orthogonally arranged microfibers with thermoresponsive and non-responsive microfibers (Figure 1). The thermoresponsive fibers, made of poly N-isopropyl acrylamide-co-glycidyl methacrylate (PNcG) with a monomer ratio of 50:1, exhibited a lower critical solution temperature (LCST) of 32°C in water.[34,35] The copolymer fibers were crosslinked post-fabrication with tetraethylenepentamine (TEPA). When immersed in water, crosslinked PNcG undergoes reversible swelling and shrinking below and above LCST, respectively. The non-responsive fibers were made of either polyvinyl cinnamate (PVCi) or poly methyl methacrylate (PMMA), which are hydrophobic and do not change their dimensions upon heating/cooling in the temperature range of the experiments.

The network was next extracted from the frame, submerged in deionized water, and placed on a heating microscope slide for environmental temperature control. The networks were initially immersed in water at a temperature of 60°C, above the LCST of PNcG, and retained their as-spun configuration. When cooled below the LCST, the PNcG fibers swelled significantly and elongated. Upon swelling, the fibers displayed two distinct shape-morphing behaviors, as was also demonstrated in our preceding investigations of homogeneous networks (see section S1 in Supporting Information).[29] The first type of morphing behavior, referred to as shape-preserving morphing (Figure 2A), was characterized by the PNcG fibers extending linearly and remaining straight throughout their swelling. Consequently, the spacing between the non-responsive fibers increased while the distances between the PNcG fibers remained constant, and the entire network expanded anisotropically along the direction of the PNcG fibers while maintaining its Cartesian structure. Upon reheating the system back above the LCST, the PNcG fibers contracted to their original size, and the network returned to its initial dimensions.

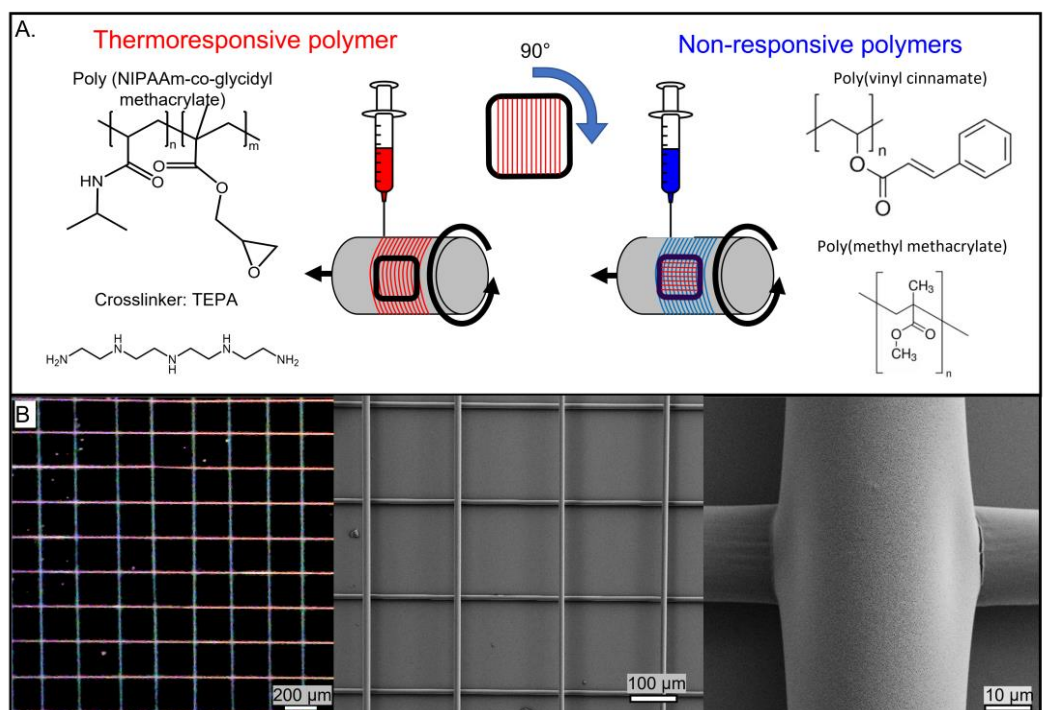

**Figure 1.** A. A schematic illustration of the fabrication process. The network is made of PNcG (red) in the horizontal direction and non-responsive polymer (blue) in the vertical direction. The chemical structure of the responsive PNcG (top left), the crosslinker TEPA (bottom left), and the non-responsive polymers PVCi (top right) and PMMA (bottom right) are depicted next to their jetting position. B. A spun network over a cover glass with PNcG fibers oriented along the horizontal axis and PVCi fibers along the vertical axis (left), and its SEM micrograph (middle). A zoom-in on a junction (node) of PNcG and PVCi in a network (right).

A second typical morphing behavior was observed wherein cooling below the LCST caused the PNcG fibers to swell, resulting in the formation of elastic instabilities and substantial buckling in the fibers (Figure 2B). Consequently, the original Cartesian network structure was disrupted and not preserved, while the non-responsive fibers and vertical node-to-node distances remained unchanged. Upon reheating, the PNcG fibers contracted, returning to their original linear configuration, and the network regained its Cartesian geometry. Remarkably, in this type of morphing, which we refer to as buckling-governed morphing, the morphing still occurs anisotropically along the responsive axis. Nonetheless, the extent of morphing was significantly lower than in the shape-preserving case, and it was strongly influenced by the buckling behavior of individual fibers. Regardless of the morphing behavior, the networks exhibited memory of the as-spun configuration and returned to it upon heating.

To better characterize the transition between the two morphing behaviors, we constructed a network architecture in which the distribution of non-responsive fibers varied along the horizontal axis while the responsive fibers were uniformly distributed along the vertical axis (Figure 3A). In this architecture, the node-to-node distance, *L*, changes along each responsive PNcG fiber, while the radius and Young modulus of the fiber remain constant. This approach enabled us to change the node-to-node distances along the responsive fibers and effectively isolate the influence of $L$ on the morphing behavior of the same fiber under controlled experimental conditions.

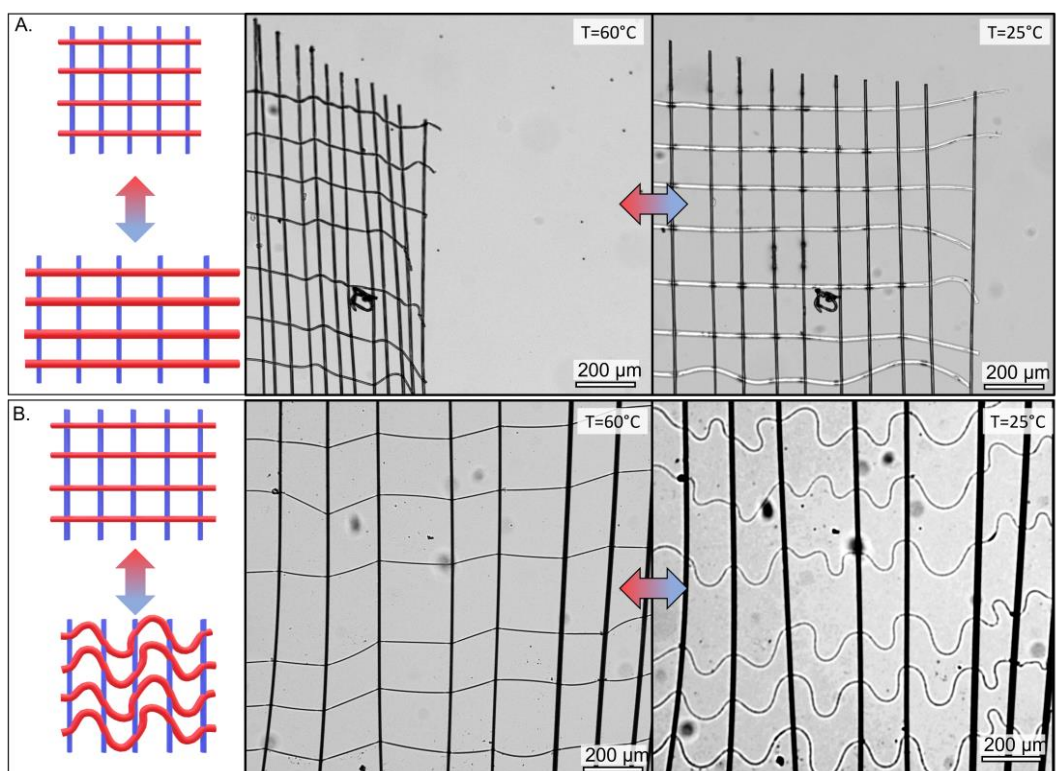

**Figure 2.** A.i. A schematic illustration of a shape-preserving behavior of a network with PNcG fibers deposited horizontally marked in red and non-responsive fibers deposited vertically marked in blue. ii. A bright-field micrograph of a representative shape-preserving network immersed in water above (left) and below (right) the LCST. B.i. A schematic illustration of a buckling-govern behavior of a network with PNcG (red) and non-responsive fibers (blue). ii. A representative buckling-governed network immersed in water above (left) and below (right) the LCST.

Figure 3B shows a time-lapse sequence of the morphing observed in a representative network of PNcG (horizontal) and PVCi (vertical) fibers, with a variable mesh size along the horizontal axis (Movie S1 in Supporting Information). At t=0, the network was immersed in water at 60°C, above the LCST. At this stage, both the responsive and non-responsive fibers remained relatively straight. As the water temperature dropped below 32°C, the responsive fibers underwent two distinct morphing behaviors depending on the mesh size dictated by the non-responsive fibers. Within dense regions with a small mesh size (indicated in pink), the responsive segments swelled and elongated linearly without buckling, preserving the network's Cartesian structure. However, in regions characterized by a larger mesh size (indicated in blue), the responsive fibers displayed significant in-plane buckling during swelling, diminishing the Cartesian structure. Notably, the non-responsive fibers remained straight through the process. The process was completely reversible, and the network retained its original shape when the temperature was reset to 60°C.

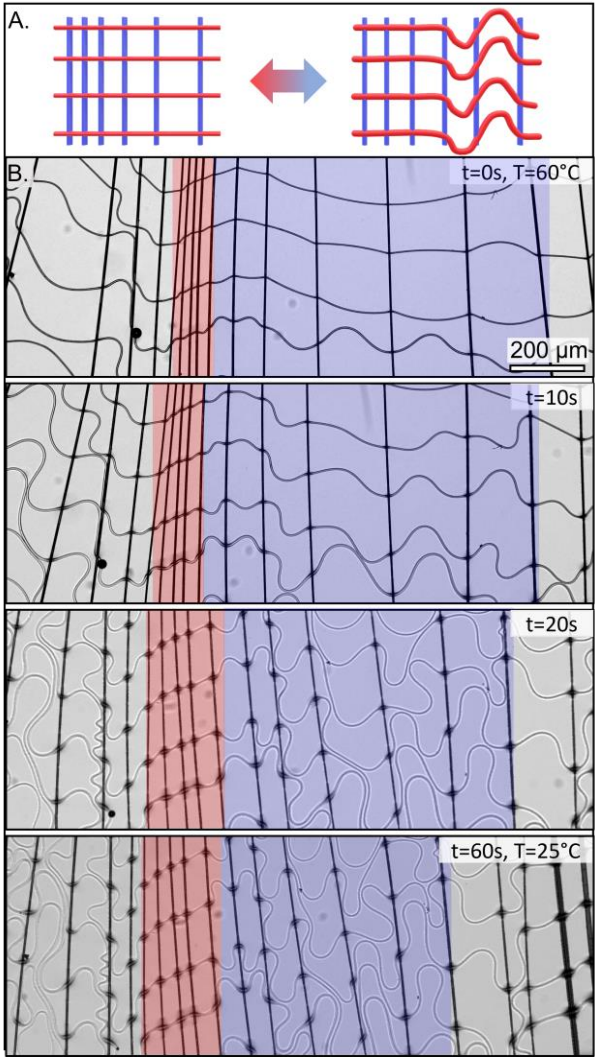

The morphing behaviors of the segments in all the networks described above, as well as previously examined homogeneous networks,[29] suggest that the change in morphing behavior can be attributed to dimensions and the mechanical properties of the filament positioned between two neighboring nodes. If the segments are approximated as Euler–Bernoulli beams, then buckling should occur if the segment is exposed to a stress that is equal to or larger than a critical Euler buckling stress, defined as

$$\sigma_{critical} = \frac{1}{4}n^2 E \pi^2 \bar{S}^2, \quad (1)$$

where $n$ is the mode order, $E$ is the Young modulus of the beam, $\bar{S}$ is the reciprocal slenderness defined as $\bar{S} = \frac{r}{L}$, $r$ is the beam's radius, and $L$ is its length. The presence of buckling in some segments suggests internal stresses caused by swelling, as no external stresses are imposed on the networks. It should be noted that Euler critical stress is calculated for a static beam, whereas the responsive fiber softens significantly as it swells and its Young modulus decreases during the process. $\bar{S}$ also changes slightly throughout the swelling process, due to different swelling along and perpendicular to the fiber's main axis. Unlike the PNcG fibers, the non-responsive fibers of both PMMA and PVCi did not exhibit significant buckling in any of the tested networks, indicating their critical buckling stress exceeds the network-induced stresses.

**Figure 3.** A. Schematic illustration of shape-morphing of nonhomogeneous network with PNcG fibers aligned horizontally (red) and non-responsive fibers aligned vertically (blue). B. Time-lapse series of the morphing of nonhomogeneous network. The pink area marks the linear extension regime, and the blue area marks the buckling-governed regime.

To further analyze the shape-morphing behavior and its dependence on the filament parameters, a quantitative indicator for the buckling extent is required. A convenient indicator is the maximum deflection of the segment from its original straight configuration, denoted as $\psi$ (Figure 4A, inset). In the case of a perfectly straight segment, $\psi$ is equal to zero, and as the degree of buckling increases, the value of $\psi$ also increases. $\psi$ can be regarded as an order parameter, indicating the extent of symmetry breaking from the straight configuration, where a value of $\psi = 40 \, \mu m$ was arbitrarily determined as the border between the straight and deflected states. Using deflection as a key parameter, we analyzed the morphing behavior of 237 segments from 34 different fibers. Figure 4A depicts a color map of the order parameter for different segments as a function of the reciprocal slenderness and of the Young modulus, calculated from the swelling ratio of the fibers using the Flory-Rehner equation (see section S2 in Supporting Information). Each segment is marked as a single point in the graph. Segments with low reciprocal slenderness exhibit significant buckling, while those with high reciprocal slenderness maintain their straight configuration. The map indicates a clear border between the two behaviors as a function of $\bar{S}$ and of the Young modulus (marked in a dashed line), and can be interpreted as a phase transition diagram.

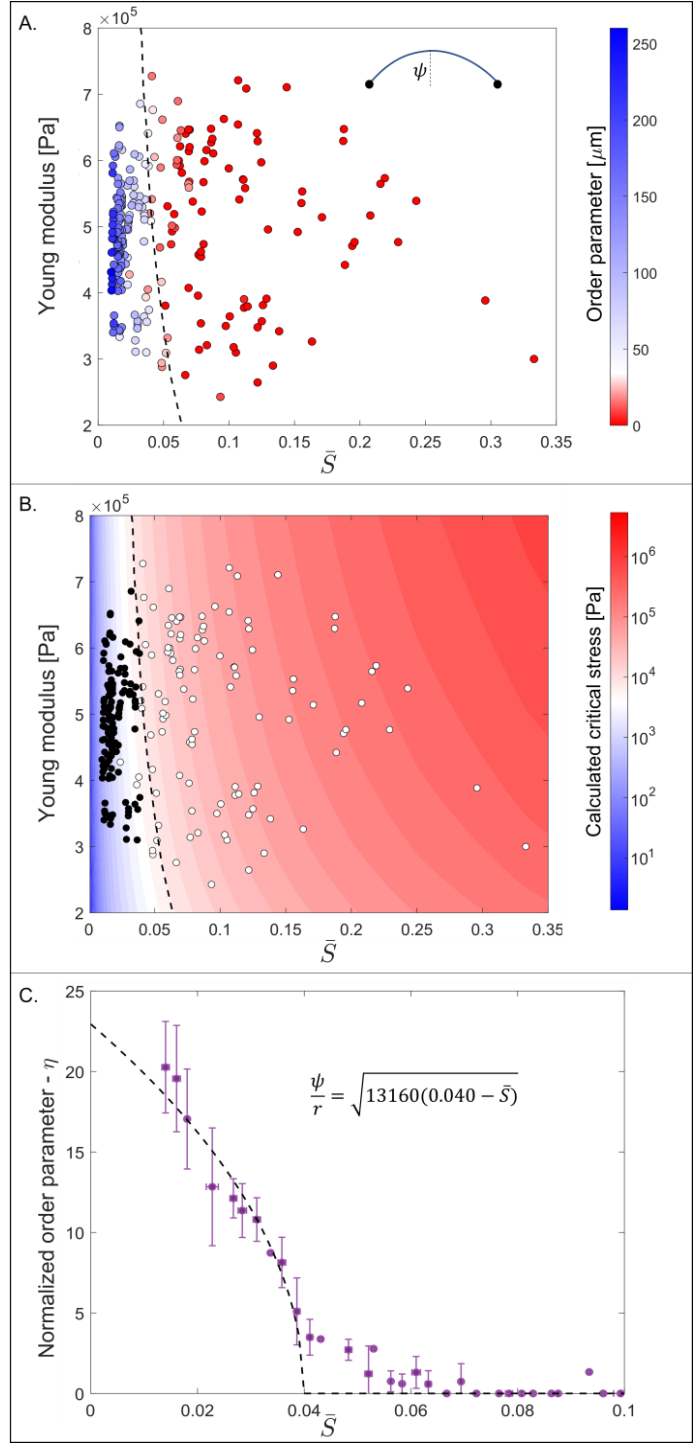

**Figure 4**. A. A color map of the order parameter as function of the reciprocal slenderness, $\bar{S}$, and of the Young modulus indicating the phase transition boundary between the two morphing behaviors. Inset shows an illustration of a buckled beam and its deflection, $\psi$. B. A map of the critical buckling stress as a function of $\bar{S}$ and the Young modulus. Circles represent the position of the examined segment on the map. Filled circles represent the buckling-governed regime and the empty circles represent the linear extension regime. The dashed black contour line represents critical buckling stresses of $10\ kPa$. C. The averaged normalized order parameter for different segments as function of $\bar{S}$ (purple dots) and the fit according to Eq. 7 (dashed line). The critical reciprocal slenderness, $\overline{S^*}$, is 0.040. Error bars represent the standard deviation.

Figure 4B presents a map of the expected Euler critical buckling stress (in logarithmic scale) as a function of the reciprocal slenderness and the Young modulus calculated according to Eq. 1. Measured segments are depicted on this map in circles, with filled circles representing segments with $\psi \geqslant 40$ µm that exhibit buckling upon swelling, and empty circles represent segments with $\psi < 40$ µm that preserve their linear morphology. The map shows that the transition between the morphing behaviors as indicated by the deflection occurs for a critical stress of about 10 kPa, thus segments whose critical stress is below that value buckle, while segments with higher critical stresses stay straight. Interestingly, this value is derived from all the segments in different networks and indicates that a similar stress of ~10kPa develops throughout the swelling process for all the networks.

The transition between morphing behaviors shown in Figures 4A and 4B calls for examining the analogy to classical phase transitions. Indeed, several works indicated the resemblance between beam buckling and Landau phase transition.[36–38] Here, we derive the analogy following the formalism suggested by Savel'ev and Nori for a beam under external force.[36] We show here only the major outcome of the derivation, which is described in detail in section S3 of the Supporting Information. For a segment of length $L$, radius $r$, and Young modulus $E$, the mechanical energy of a segment, $F$, under an externally applied stress $\sigma^*$, is given by:

$$F = \int_0^L dl \left\{ \frac{EI(y'')^2}{2(1-(y')^2)} + \pi r^2 \sigma^* \left( \sqrt{(1-(y')^2)} - 1 \right) \right\}, \qquad (2)$$

where $I$ is the area moment of inertia, which for a cylinder is given by $I = \frac{\pi r^4}{4}$, and $y$ is the contour of the beam with respect to the beam axis coordinate, $l$. All the derivatives are taken with respect to the beam axis coordinate, $l$. The first term of the integral provides the energy due to buckling, while the second term is the mechanical work performed by the force. For the sake of simplicity, we assume no moments are exerted on the segments at the initial unswollen state. This is an approximation for the experimental system, in which moments may be formed due to small defects in the network fabrication or to external factors, yet as can be seen in Figures 2 and 3, and in Movie S1, these mostly affect the segments in the edge of the network yet are relatively small and negligible in controlling the morphing process of the inner segments.

If we assume the segment is regarded as an Euler beam, the solution for the beam contour obtained from Euler analysis is:

$$y(l) = \psi \sin\left(\frac{n\pi l}{L}\right), \quad (3)$$

where $\psi$ is the deflection amplitude. By introducing the solution of the beam contour into Eq. 2, expanding it with respect to $\psi$, and integrating, the energy is expressed as:

$$F = \frac{\pi^3 n^2 r^2}{4L}\left(\frac{r^2}{4}\frac{n^2 E \pi^2}{L^2} - \sigma^*\right)\psi^2 + \frac{\pi^5 n^4 r^2}{64 L^3}\left(4\frac{r^2}{4}\frac{n^2 E \pi^2}{L^2} - 3\sigma^*\right)\psi^4 + O(\psi^6). \quad (4)$$

For all our samples, the external stress applied to the segments, $\sigma^*$, as well as the Young modulus, $E$, are very similar and can be taken as constants. This implied that for all the systems there exists a reciprocal slenderness, $\overline{S^*}$, for which the stress $\sigma^* = \frac{n^2 E \pi^2}{4}\overline{S^*}^2$ will cause buckling. Substituting this expression into Eq. 4 yields:

$$F = \frac{\pi^5 n^4 E r^2}{16L}(\overline{S} + \overline{S^*})(\overline{S} - \overline{S^*})\psi^2 + \frac{\pi^7 n^6 E r^2}{256 L^3}\left(4\overline{S}^2 - 3\overline{S^*}^2\right)\psi^4 + O(\psi^6). \quad (5)$$

Eq. 5 is written as Landau-type energy expansion, where the reciprocal slenderness, $\overline{S}$, is the driving parameter corresponding to temperature in the standard expression, and the amplitude, $\psi$, acts as the order parameter.

The order parameter that minimizes the energy can be found by solving the equation $\frac{dF}{d\psi} = 0$, which yields that the relation between the order parameter and the driving parameter in proximity to the critical reciprocal slenderness ($\overline{S} \to \overline{S^*}$) is:

$$\psi = r\sqrt{-\frac{16}{\pi^2 n^2 \overline{S^*}^3}(\overline{S} - \overline{S^*})} \propto \sqrt{(\overline{S} - \overline{S^*})}. \quad (6)$$

However, this expression depends on the radius of the fiber. To attain a general expression for segments of different radii, the amplitude can be divided by $r$, generating a normalized order parameter:

$$\eta \equiv \frac{\psi}{r} \simeq \begin{cases} 0, & \overline{S} > \overline{S^*} \\ \sqrt{\alpha(\overline{S^*} - \overline{S})}, & \overline{S} < \overline{S^*} \end{cases}, \quad (7)$$

where $\alpha = \frac{16}{\pi^2 n^2 \overline{S^*}^4}$ is a positive constant.

Figure 4C shows the averaged normalized order parameter for different segments as a function of the reciprocal slenderness, $\overline{S}$, (purple dots) and the fit to these experimental results according to Eq. 7 (dashed black line). There is a high level of agreement between the values obtained from the experimental results and the theoretical expression, indicating a critical slenderness of $\overline{S^*} = 0.040$.

The change in the morphing behavior occurs on the filament level at the microscale and propagates through the hierarchical network structures to induce morphological changes at the macroscale. Since the network structure is primarily defined by its nodes, the dimensions of the network are determined by the distance between adjacent nodes along the network. Thus, in regions where the filaments undergo shape-preserving morphing and remain straight, as depicted in pink in Figure 3B, the distance between the nearest neighbor nodes is equal to the length of the connecting filament. Upon swelling of the filaments, the entire network expands, and the local swelling ratio of the network is comparable to that of the single filaments. This outcome is expected because the change in the filament length induces compressive stress that is distributed uniformly along the main axis of the fiber, resulting in the motion of the nodes along this direction, which leads to the swelling of the network.

The relationship between the morphing behavior of the network and the filament morphing in the buckling-governed regime is more intricate. Buckling leads to a considerable modification in stress distribution within the filament. The compressive stress in the filament becomes highly concentrated near the points of maximum deflection, whereas the stress in the remaining regions of the filament and near the nodes, reduces significantly. Under these conditions, it is expected that the distance between the nodes will be smaller than the longitudinal swelling of the filaments, and the network will exhibit significantly less swelling in comparison to the shape-preserving scenario. To examine this, the ratio between the filament length, $L_s$, and the node-to-node distance, $\xi_s$, in the swollen state was measured for 237 segments and the average value normalized order parameter, $\eta$, of the segments is portrayed in Figure 5A (errors represent the standard deviation). This ratio equals 1 for linearly extending segments and increases with the normalized order parameter, indicating that the length of the filaments increases more rapidly than the node-to-node distance.

Nevertheless, upon examining the dimensions of the regions in the networks that exhibit buckling-governed morphing, e.g., the area that is depicted in Figure 3B in blue, a surprising observation arises – not only that the network in this region does not elongate upon swelling of the filaments, but it significantly contracts in length to about 1.5 times less than in the unswollen state. To examine this phenomenon, we compared the distance between the adjacent nodes in the swollen and unswollen networks ($\xi_s$ and $\xi_{us}$ respectively) for the same segments that were measured in Figure 5A. Figure 5B illustrates the average ratio $\xi_s/\xi_{us}$ along the PNcG fibers as a function of the normalized order parameter, $\eta$ (errors represent the standard deviation). For low order parameters, the ratio is greater than 1, indicating swelling of the network. As the order parameter increases the ratio decreases, eventually going below 1 for $\eta > 17$. The contraction of the network at high order parameters is counterintuitive and reminiscent of negative compressibility. In most materials, applying a longitudinal force leads to a longitudinal deformation. However, in materials with negative compressibility, the longitudinal deformation opposes the longitudinal applied force. Hence the system contracts under tension and dilates under pressure. Here, upon longitudinal swelling of the fibers, the network contracts. A plausible explanation for the observed behavior could be attributed to the spontaneous shearing of the network in a direction perpendicular to the axis of the PNcG fibers. Figure S1 in section S4 the Supporting Information displays the node-to-node pathways of the PNcG fibers atop the network shown in Figure 3B. In the buckling-governed regimes, during the progression of the morphing process, several non-responsive fibers (and their associated nodes) exhibit considerable vertical translation along their primary axis. This motion can be interpreted as a shearing of the network. Due to this shearing, the responsive fiber segments that are linked to the moving fiber are pulled, resulting in the reduction of their buckling. This reduction is energetically favorable since it decreases the buckling energy without increasing the stretching energy of the segment. However, the pulling reaction of the responsive fibers leads to the pulling of the non-responsive fibers in the opposite direction, resulting in a decrease in the distance between them and in the contraction of the network in a scissors mechanism. Such shearing requires buckling of the fibers, and hence cannot occur in the shape-preserving regions where the critical buckling load is high.

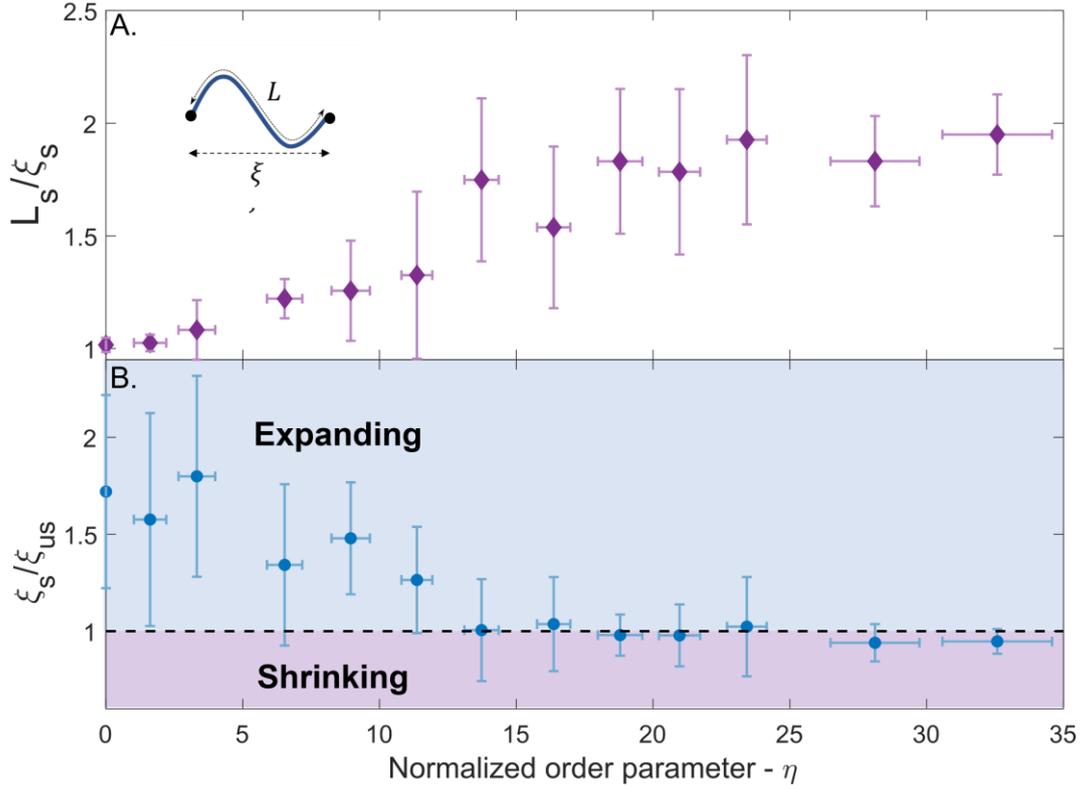

**Figure 5.** A. The average ratio of the filament length ($L_s$) to the node-to-node distance ($\xi_s$) in the swollen state as function of the normalized order parameter, $\eta$. Inset: an illustration of a buckled segment showing the corresponding lengths. B. The average ratio of the node-to-node distance in the swollen and shrunk states ($\xi_s$ and $\xi_{us}$ respectively) as function of the normalized order parameter. As the buckling is increased, the ratio decreases, and above an order parameter of ~17 the ratio decreases below 1, indicating shrinking of the network. In both graphs, the error bars indicate the standard deviation.

All the aforementioned findings have been examined within the confines of relatively small networks characterized by an area of approximately a square millimeter. This prompts the inquiry as to whether controllability over directional deformations and structural changes can be extrapolated to macroscale networks. To investigate this prospect, we meticulously assembled macroscale anisotropic networks measuring 4 centimeters along the inert axis and 2 centimeters along the responsive axis—dimensions spanning three to four orders of magnitude larger than the fiber diameter and length, respectively. Figure 6 and Movie S2 show the reversible shape-morphing of a macro-network with an average swollen segment length of 450 μ$m$ and radius of $r = 25$ μ$m$, corresponding to an average reciprocal slenderness of $\bar{S} = 0.06$. Such a network is anticipated to be shape-preserving and extend inhomogeneously as it swells. Indeed, it manifests a swelling ratio of 1.73, a value consistent with the swelling ratios observed in individual microscale segments, as illustrated in Figure 5B. In macro-networks with lower $\bar{S}$, in which the morphing is governed by fiber buckling, no swelling was observed. Thus, the microscale morphing is maintained also when increasing the size of the networks to centimeters. Notably, larger networks are susceptible to the presence of defects including loose responsive fibers, which can be easily identified by their constant buckled state in both swollen and unswollen states. Akin buckling behavior is also observed at the periphery of the network in the free ends of the fibers. In addition, larger networks are submerged within a more substantial volume of water, resulting in extended heating and cooling cycles and correspondingly protracted morphing durations compared to their micro-network counterparts.

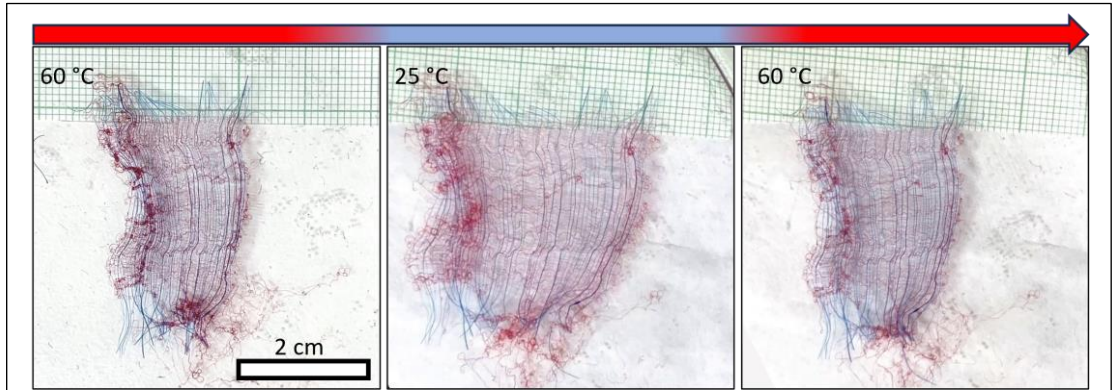

**Figure 6:** A typical shape-preserving macroscale network exhibiting swelling and shrinking upon temperature changes, with network width going from ~2cm in the shrunk state to ~3.5cm in the swollen state, while the length remains constant (~4 cm). Responsive fibers are dyed in red, and inert fibers in blue. Loose responsive fibers can be identified by their permanent buckled state.

In conclusion, we demonstrated the fabrication and characterization of microscale and macroscale inhomogeneous shape-morphing networks composed of thermoresponsive and non-responsive microfibers. Through a hierarchical construction of the networks and precise positioning of fibers with different responses, we successfully demonstrated the ability to induce directional actuation of the networks. The shape-morphing of individual segments within the network can exhibit one of two distinct behaviors: buckling or linear extension, depending on the radius and length of the segment and the Young modulus of the polymer. The transition between these two morphing behaviors follows a second-order athermal Landau-like phase transition, where the buckling amplitude serves as the order parameter, while the reciprocal slenderness acts as the effective temperature. The disparities in morphing behaviors upon swelling result in significant differences in the macroscale morphing of the network itself, where regions of linear extension preserve their Cartesian structure and exhibit swelling, while buckling-governed regions experience shrinking compared to the unswollen state and exhibit internal negative compressibility. The hierarchical control of the macroscale morphing through adjustments of the microscale components offers new insights regarding the physical behavior of shape-morphing structures constructed of soft microscale filaments. It also introduces a novel approach for modulating the morphing and for programming responsive systems with a mesoscale resolution, thereby enabling the development of innovative applications in tissue engineering, soft robotics, and micro-actuation.

**Experimental Methods**

**Materials**: N-isopropyl acrylamide (NIPAAm) 99%, glycidyl methacrylate (GMA), and 2,2′-Azobis (2-methyl propionitrile) (AIBN) were purchased from Alfa Aesar. Toluene (anhydrous, 99.8%), dimethylformamide (DMF), tetrahydrofuran (THF), chloroform, dichloromethane (DCM), hexane, tetraethylenepentamine (TEPA), polyvinyl cinnamate (PVCi), and poly methyl methacrylate (PMMA) were purchased from Sigma-Aldrich. Diethyl ether was purchased from BioLab. All the materials were used as bought, without further purification.

**Copolymer Synthesis:** The copolymer poly (N-isopropyl acrylamide-co-glycidyl methacrylate) PNcG was synthesized as was previously reported.[29] See section S5 in the Supporting Information.

**Network Fabrication:** The 2D Cartesian networks were fabricated via the dry-spinning process, using thermoresponsive fibers in one axis, and non-responsive in the orthogonal axis. For the fabrication of thermoresponsive fibers, a 0.50-0.75 g mL$^{-1}$ solution was prepared by dissolving PNcG in a mixture of chloroform and DMF (1:1 v/v). 6% of TEPA, was added before the jetting process. PVCi and PMMA were chosen to fabricate non-responsive fibers. 0.7 g of PVCi was dissolved in 1 mL of a mixture of THF and DMF (6:4 v/v). 0.3 g of PMMA was dissolved in 1 mL of a mixture of DMF and chloroform (1:1 v/v).

For constructing Cartesian networks, the PNcG solution was dispensed via a metallic needle at a flow rate of 0.006-0.020 mL h$^{-1}$. The rotating speed of the drum was 14-30 mm s$^{-1}$. A linear stage velocity of 0.05-0.35 mm s$^{-1}$ was set according to the rotating velocity to obtain the desired gaps. The non-responsive fibers solutions were dispensed via a metallic gauge needle at a flow rate of 0.014 mL h$^{-1}$ for the PMMA solution and 0.140 mL h$^{-1}$ for the PVCi solution. The rotating speed of the drum was 7 mm s$^{-1}$ for the PMMA solution, and 14 mm s$^{-1}$ for the PVCi solution. To create a density gradient, the stage velocity was varied from 0.12 to 0.50 mm s$^{-1}$ for PMMA fibers, and from 0.17 to 2.16 mm s$^{-1}$ for the PVCi fibers. Finally, the networks were cured post-fabrication at 70°C overnight in an oven. See section S5 in the Supporting Information.

**Instrumentation**: Scanning electron microscopy (SEM) was performed using a Zeiss GeminiSEM 300, in a high vacuum, WD ~7 mm, 3 kV. All images and videos were taken with Olympus, IX73 microscope equipped with a heating glass slide (LCI, CU-301). The spinning setup included a syringe pump (New Era), a linear motion stage (ILS-200LM, Newport), an eight-axis universal controller (XPS-D8, Newport), and a rotating drum collector.

**Notes**
The authors declare no competing financial interest.

**Funding:**

This research was funded by generous support from the Israel Science Foundation (grant no. 1682/22) and the Army Research Office (Grant Number W911NF-23-1-0257). The views and conclusions contained in this document are those of the authors and should not be interpreted as representing the official policies, either expressed or implied, of the Army Research Office or the U.S. Government. The U.S. Government is authorized to reproduce and distribute reprints for Government purposes notwithstanding any copyright notation herein.

**Acknowledgments**

S.Z.S. and N.E.-P. acknowledge the generous support of The Shulamit Aloni Scholarship for Advancing Women in Exact Science and Engineering, provided by The Ministry of Science Technology, Israel. The authors acknowledge the Chaoul Center for Nanoscale Systems of Tel Aviv University for the use of instruments and staff assistance, and the Mechanical Workshop for Research and Development, School of Chemistry, Tel Aviv University, for their help in constructing the fabrication devices.

**Supporting Information Available:**

Uniform Cartesian network analysis, Flory-Rehner equation, Using phase transition formalism for describing buckling of a beam, Figure of shearing within buckling-governed morphing regimes, Detailed experimental section, Movie S1 of a representative network of PNcG and PVCi fibers, with a variable mesh size along the horizontal axis, Movie S2 of a macroscale network exhibiting swelling and shrinking upon temperature changes.

**References**

(1) Wang, J.; Gao, D.; Lee, P. S. Recent Progress in Artificial Muscles for Interactive Soft Robotics. *Adv. Mater.* **2021**, *33*, 2003088.

(2) Park, N.; Kim, J. Hydrogel-Based Artificial Muscles: Overview and Recent Progress. *Adv. Intell. Syst.* **2020**, *2*, 1900135.

(3) Kim, H.; Ahn, S. kyun; Mackie, D. M.; Kwon, J.; Kim, S. H.; Choi, C.; Moon, Y. H.; Lee, H. B.; Ko, S. H. Shape Morphing Smart 3D Actuator Materials for Micro Soft Robot. *Mater. Today* **2020**, *41*, 243–269.

(4) Liu, X.; Gao, M.; Chen, J.; Guo, S.; Zhu, W.; Bai, L.; Zhai, W.; Du, H.; Wu, H.; Yan, C.; Shi, Y.; Gu, J.; Qi, H. J.; Zhou, K. Recent Advances in Stimuli-Responsive Shape-Morphing Hydrogels. *Adv. Funct. Mater.* **2022**, *32*, 2203323.

(5) Wong, J.; Basu, A.; Wende, M.; Boechler, N.; Nelson, A. Mechano-Activated Objects with Multidirectional Shape Morphing Programmed via 3D Printing. *ACS Appl. Polym. Mater.* **2020**, *2*, 2504–2508.

(6) Yiming, B.; Liu, T.; Nian, G.; Han, Z.; Jia, Z.; Qu, S. Mechanics-Guided Design of Shape-Morphing Composite Sheets with Hard and Soft Materials. *Extrem. Mech. Lett.* **2020**, *35*, 100643.

(7) Wu, Z. L.; Moshe, M.; Greener, J.; Therien-aubin, H.; Nie, Z.; Sharon, E.; Kumacheva, E. Three-Dimensional Shape Transformations of Hydrogel Sheets Induced by Small-Scale Modulation of Internal Stresses. *Nat. Commun.* **2013**, *4*, 1586.

(8) Siéfert, E.; Reyssat, E.; Bico, J.; Roman, B. Bio-Inspired Pneumatic Shape-Morphing Elastomers. *Nat. Mater.* **2019**, *18*, 24–28.

(9) Meza, L. R.; Das, S.; Greer, J. R. Strong, Lightweight, and Recoverable Three-Dimensional Ceramic Nanolattices. *Science.* **2014**, *345*, 1322–1326.

(10) Xia, X.; Spadaccini, C. M.; Greer, J. R. Responsive Materials Architected in Space and Time. *Nat. Rev. Mater.* **2022**, *7*, 683–701.

(11) Kucewicz, M.; Baranowski, P.; Małachowski, J.; Popławski, A.; Płatek, P. Modelling, and Characterization of 3D Printed Cellular Structures. *Mater. Des.* **2018**, *142*, 177–189.

(12) Yu, X.; Zhou, J.; Liang, H.; Jiang, Z.; Wu, L. Mechanical Metamaterials Associated with Stiffness, Rigidity, and Compressibility: A Brief Review. *Prog. Mater. Sci.* **2018**, *94*, 114–173.

(13) Dudek, K. K.; Martínez, J. A. I.; Ulliac, G.; Kadic, M. Micro-Scale Auxetic Hierarchical Mechanical Metamaterials for Shape Morphing. *Adv. Mater.* **2022**, *34*, 2110115.

(14) Nojoomi, A.; Jeon, J.; Yum, K. 2D Material Programming for 3D Shaping. *Nat. Commun.* **2021**, *12*, 603.

(15) Klein, Y.; Efrati, E.; Sharon, E. Shaping of Elastic Sheets by Prescription of Non-Euclidean Metrics. *Science.* **2007**, *315*, 1116–1120.

(16) Dong, L.; Wang, J.; Wang, D. Modeling and Design of Three-Dimensional Voxel Printed Lattice Metamaterials. *Addit. Manuf.* **2023**, *69*, 103532.

(17) Fan, X.; Deng, C.; Gao, H.; Jiao, B.; Liu, Y.; Chen, F.; Deng, L.; Xiong, W. 3D Printing of Nanowrinkled Architectures via Laser Direct Assembly. *Sci. Adv.* **2022**, *8*, 1–11.

(18) Mueller, J.; Lewis, J. A.; Bertoldi, K. Architected Multimaterial Lattices with Thermally Programmable Mechanical Response. *Adv. Funct. Mater.* **2022**, *32*, 2105128.

(19) Ashby, M. F. Mechanical Properties of Cellular Solids. *Metall. Trans. A, Phys. Metall. Mater. Sci.* **1983**, *14*, 1755–1769.

(20) Gladman, A. S.; Elisabetta A. Matsumoto; Ralph G. Nuzzo; Mahadevan, L.; Lewis, J. A. Biomimetic 4D Printing. *Nat. Mater.* **2016**, *15*, 413–419.

(21) Boley, J. W.; Van Rees, W. M.; Lissandrello, C.; Horenstein, M. N.; Truby, R. L.; Kotikian, A.; Lewis, J. A.; Mahadevan, L. Shape-Shifting Structured Lattices via Multimaterial 4D Printing. *Proc. Natl. Acad. Sci. U. S. A.* **2019**, *116*, 20856–20862.

(22) Gregg, A.; De Volder, M. F. L.; Baumberg, J. J. Light-Actuated Anisotropic Microactuators from CNT/Hydrogel Nanocomposites. *Adv. Opt. Mater.* **2022**, *10*, 2200180.

(23) Li, S.; Deng, B.; Grinthal, A.; Schneider-Yamamura, A.; Kang, J.; Martens, R. S.; Zhang, C. T.; Li, J.; Yu, S.; Bertoldi, K.; Aizenberg, J. Liquid-Induced Topological Transformations of Cellular Microstructures. *Nature* **2021**, *592*, 386–391.

(24) Cheng, X.; Fan, Z.; Yao, S.; Jin, T.; Lv, Z.; Lan, Y.; Bo, R.; Chen, Y.; Zhang, F.; Shen, Z.; Wan, H.; Huang, Y.; Zhang, Y. Programming 3D Curved Mesosurfaces Using Microlattice Designs. *Science* **2023**, *379*, 1225–1232.

(25) Liu, L.; Jiang, S.; Sun, Y.; Agarwal, S. Giving Direction to Motion and Surface with Ultra-Fast Speed Using Oriented Hydrogel Fibers. *Adv. Funct. Mater.* **2016**, *26*, 1021–1027.

(26) Jiang, S.; Liu, F.; Lerch, A.; Ionov, L.; Agarwal, S. Unusual and Superfast Temperature-Triggered Actuators. *Adv. Mater.* **2015**, *27*, 4865–4870.

(27) He, Q.; Wang, Z.; Wang, Y.; Wang, Z.; Li, C.; Annapooranan, R.; Zeng, J.; Chen, R.; Cai, S. Electrospun Liquid Crystal Elastomer Microfiber Actuator. *Sci. Robot.* **2021**, *6*, eabi9704.

(28) Moon, S.; Jones, M. S.; Seo, E.; Lee, J.; Lahann, L.; Jordahl, J. H.; Lee, K. J.; Lahann, J. 3D Jet Writing of Mechanically Actuated Tandem Scaffolds. *Sci. Adv.* **2021**, *7*, eabf5289.

(29) Ziv Sharabani, S.; Edelstein-Pardo, N.; Molco, M.; Bachar Schwartz, N.; Morami, M.; Sivan, A.; Gendelman Rom, Y.; Evental, R.; Flaxer, E.; Sitt, A. Messy or Ordered? Multiscale Mechanics Dictates Shape-Morphing of 2D Networks Hierarchically Assembled of Responsive Microfibers. *Adv. Funct. Mater.* **2022**, *32*, 2111471.

(30) Jordahl, J. H.; Solorio, L.; Sun, H.; Ramcharan, S.; Teeple, C. B.; Haley, H. R.; Lee, K. J.; Eyster, T. W.; Luker, G. D.; Krebsbach, P. H.; Lahann, J. 3D Jet Writing: Functional Microtissues Based on Tessellated Scaffold Architectures. *Adv. Mater.* **2018**, *30*, 1707196.

(31) Nahm, D.; Weigl, F.; Schaefer, N.; Sancho, A.; Frank, A.; Groll, J.; Villmann, C.; Schmidt, H. W.; Dalton, P. D.; Luxenhofer, R. A Versatile Biomaterial Ink Platform for the Melt Electrowriting of Chemically-Crosslinked Hydrogels. *Mater. Horizons* **2020**, *7*, 928–933.

(32) Wu, D. J.; Vonk, N. H.; Lamers, B. A. G.; Castilho, M.; Malda, J.; Hoefnagels, J. P. M.; Dankers, P. Y. W. Anisotropic Hygro-Expansion in Hydrogel Fibers Owing to Uniting 3D Electrowriting and Supramolecular Polymer Assembly. *Eur. Polym. J.* **2020**, *141*, 110099.

(33) Javadzadeh, M.; del Barrio, J.; Sánchez-Somolinos, C. Melt Electrowriting of Liquid Crystal Elastomer Scaffolds with Programmed Mechanical Response. *Adv. Mater.* **2023**, *2209244*.

(34) Schild, H. G. Poly(N-Isopropylacrylamide): Experiment, Theory and Application. *Prog. Polym. Sci.* **1992**, *17*, 163–249.

(35) Halperin, A.; Kröger, M.; Winnik, F. M. Poly(N-Isopropylacrylamide) Phase Diagrams: Fifty Years of Research. *Angew. Chemie* **2015**, *54*, 15342–15367.

(36) Savel'ev, S.; Nori, F. Magnetic and Mechanical Buckling: Modified Landau Theory Approach to Study Phase Transitions in Micromagnetic Disks and Compressed Rods. *Phys. Rev. B - Condens. Matter Mater. Phys.* **2004**, *70*, 1–19.

(37) Bobnar, J.; Susman, K.; Parsegian, V. A.; Rand, P. R.; Čepič, M.; Podgornik, R. Euler Strut: A Mechanical Analogy for Dynamics in the Vicinity of a Critical Point. *Eur. J. Phys.* **2011**, *32*, 1007–1018.

(38) Margaretta, D. O.; Amalia, N.; Utami, F. D.; Viridi, S.; Abdullah, M. Second-Order Phase Transition and Universality of Self-Buckled Elastic Slender Columns. *J. Taibah Univ. Sci.* **2019**, *13*, 1128–1136.